# Security of high speed quantum key distribution with finite detector dead time

**Viacheslav Burenkov[1,5], Bing Qi[2,3], Ben Fortescue[4] and Hoi-Kwong Lo[1,2,3]**

[1] Department of Physics, University of Toronto, Toronto, Ontario, M5S 1A7, Canada
[2] Center for Quantum Information and Quantum Control, University of Toronto, Toronto, Ontario, Canada
[3] Department of Electrical and Computer Engineering, University of Toronto, Ontario, M5S 3G4, Canada
[4] Institute for Quantum Information Science, University of Calgary, Alberta, T2N 1N4, Canada

Email: viacheslav.burenkov@utoronto.ca

**Abstract.** The security of a high speed quantum key distribution system with finite detector dead time $\tau$ is analyzed. When the transmission rate becomes higher than the maximum count rate of the individual detectors ($1/\tau$), security issues affect the algorithm for sifting bits. Analytical calculations and numerical simulations of the Bennett-Brassard BB84 protocol are performed. We study Rogers et al.'s protocol (introduced in "Detector dead-time effects and paralyzability in high-speed quantum key distribution," *New J. Phys.* **9** (2007) 319) in the presence of an active eavesdropper Eve who has the power to perform an intercept-resend attack. It is shown that Rogers et al.'s protocol is no longer secure. More specifically, Eve can induce a basis-dependent detection efficiency at the receiver's end. Modified key sifting schemes that are secure in the presence of dead time and an active eavesdropper are then introduced. We analyze and compare these secure sifting schemes for this active Eve scenario, and calculate and simulate their key generation rate. It is shown that the maximum key generation rate is $1/(2\tau)$ for passive basis selection, and $1/\tau$ for active basis selection. The security analysis for finite detector dead time is also extended to the decoy state BB84 protocol.

**Contents**

[5] Author to whom any correspondence should be addressed.

## 1. Introduction

Quantum key distribution (QKD) [1,2] can be used to expand a secret key (random bit string) between two distant parties, Alice and Bob. The security of the key is guaranteed by the laws of quantum mechanics; the process of measuring a quantum system generally disturbs it, thus allowing Alice and Bob to detect the presence of an eavesdropper, Eve. For a review of QKD, see [3-5].

QKD is unconditionally secure [6-8] and BB84 [1] is a commonly used QKD protocol. It can be used in its original form or supplemented by the use of decoy states [9-11] for improved performance.

As the length of the secret key needs to be as long as the message for secure one-time-pad encryption, the secret key generation rate is a crucial figure of merit. As such, there has been a lot of recent progress in experimental high speed QKD [12,13]. It is generally true that increasing the transmission rate increases the secret key generation rate. However, since most realistic single photon detectors have a property called dead time – the time interval right after a detection, during which a detector recovers and cannot detect another incoming photon – certain security assumptions may be violated if transmission rates are increased inattentively. In this paper we consider the security of a high speed QKD system with finite detector dead time, in the regime where the transmission rate is so high that photons can arrive at Bob's detectors while one or more detectors are still recovering from previous detection events. This work builds on the earlier work by D. Rogers et al. [14]. Monte-Carlo simulations of the BB84 protocol were performed to extend the security analysis to include an active Eve capable of interfering with the signals.

In section 2, we describe the model and a sifting algorithm proposed by Rogers et al., for secure operation in the passive Eve scenario. In section 3, we show that this algorithm is no longer secure when Eve is able to perform an intercept-resend attack. In section 4, we analyze sifting schemes that are secure in this active Eve scenario and compare the sifted bit rate of these algorithms. Section 5 extends the sifted key rate analysis to the decoy-state BB84 protocol. Section 6 contains our concluding remarks.

## 2. System model and Rogers' algorithm

We consider polarization encoded BB84 protocol, with a passive polarization detection scheme [3] shown in figure 1. Notice that there are two bases:

1) rectilinear (consisting of vertically and horizontally polarized photons) and
2) diagonal (consisting of 45-degree and 135-degree polarized photons).

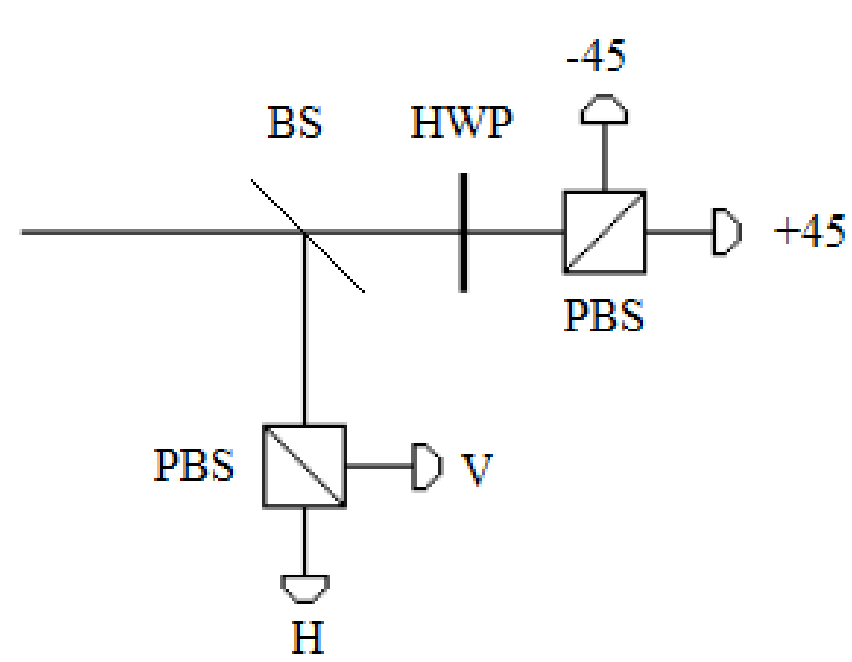

**Figure 1.** Passive polarization detection optics. A 50-50 beam splitter (BS) performs basis selection and the combination of half waveplate (HWP) and polarizing beam splitters (PBS) perform the polarization measurement in two bases: basis 1 (V-H) and basis 2 (-45˚, +45˚).

The idealized gain $\eta$ is given by the overall transmission and detection efficiency [15]:

$$\eta = t_{AB}\eta_{Bob} = t_{AB}t_{Bob}\eta_{D}, \tag{1}$$

where

$t_{AB}$ = channel transmittance
$\eta_{Bob}$ = efficiency of Bob's system
$t_{Bob}$ = Bob's internal transmittance
$\eta_{D}$ = detection efficiency.

The channel transmittance can be expressed as:

$$t_{AB} = 10^{-\frac{\alpha l}{10}}, \tag{2}$$

where $\alpha$ is the loss coefficient in dB/km and $l$ is the channel distance in km.

The list of assumptions about the QKD system is provided below.

Except for Sections 5 and 6, the source is assumed to be a perfect single photon source. A number of randomly and uniformly chosen signals sent by Alice are erased to model the loss in the quantum channel. Fibre loss $\alpha$ is about 0.2 dB/km at the telecom wavelength of 1550 nm. As such, a typical loss for a 100 km length of fibre in current experiments is of the order of 20 dB.

The detector model is as follows. We assume detectors have a dead time $\tau$. In Si-APDs, during the dead time, the bias voltage across the p-n junction is below the breakdown threshold and as such, another photon cannot be detected [16]. This limits their counting rate to $1/\tau$. In our simplified model, we assume that an active detector will detect an incoming signal with some maximum constant detection efficiency $\eta_D$, which drops instantaneously to 0 after a detection event, and undergoes an instantaneous transition back to $\eta_D$ after the dead time $\tau$ (see figure 2).

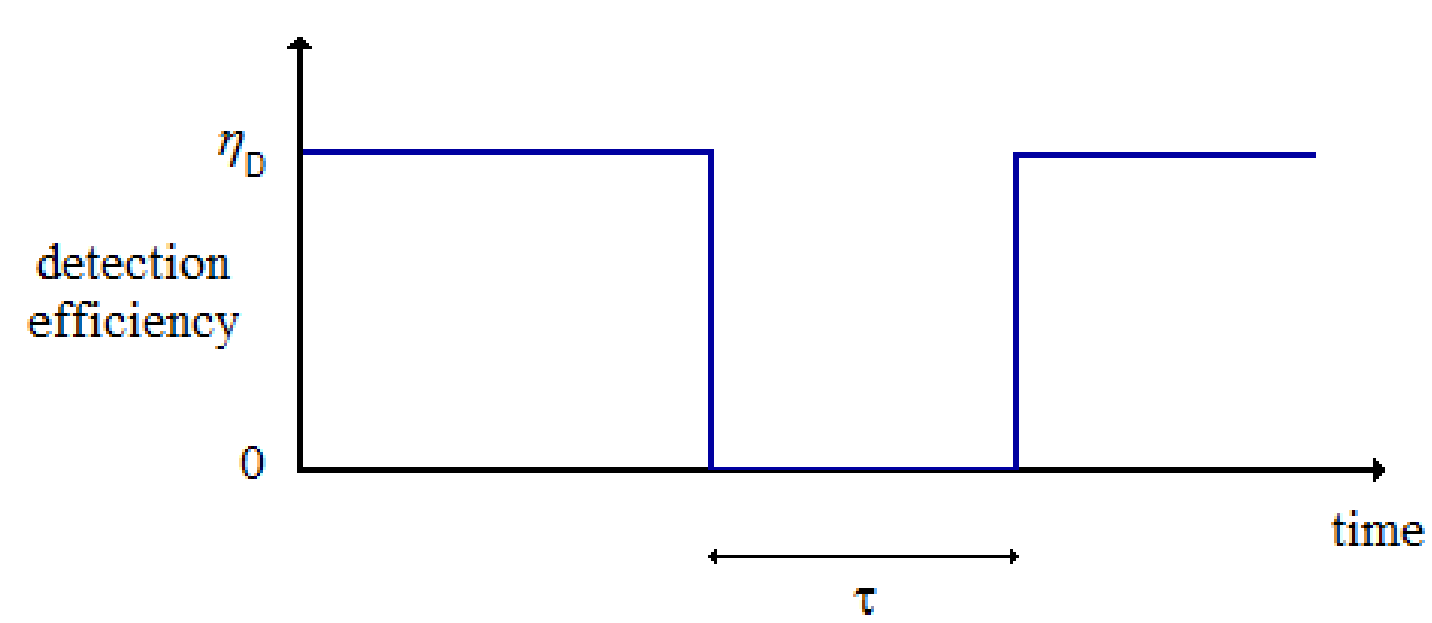

**Figure 2.** Detection dead time model. The detector undergoes an instantaneous transition from some constant maximum value $\eta_D$ to 0% efficiency upon being hit by a photon and instantly back again after a dead time $\tau$.

For Silicon SPADs the typical dead time is of the order of 100 ns [16], which is the value used in our simulations. All four detectors are assumed to have the same dead time. We further assume that a photon that strikes the detector while it is recovering does not extend the recovery time, nor has any effect whatsoever [17]. This makes individual detectors non-paralyzable systems [18]. Finally, it is assumed that there are no dark counts (except for sections 5 and 6) and the channel is noiseless.

In this section, it is assumed that Eve is passive. In other words, she does not interfere with the quantum signals but she can 'listen' to the classical channel so that she has full information about the classical transmission from Bob to Alice (bases used and time of detections) and Alice to Bob (which detections to sift).

Generally, as the transmission rate increases, the sifted key rate increases. However, since the detectors have a finite dead time $\tau$, there comes a point where the transmission rate $\rho$ (in terms of number of transmissions per second) is so high that it exceeds the maximum counting rate of the individual detectors ($1/\tau$), so that photons can arrive at Bob while one or more of his detectors are recovering from previous detection events. If two detection events occur in the same basis within one dead time window, they necessarily correspond to two *different* bits [19]. This leads to correlations in the sifted bit string, which is an obvious security flaw. Each 'closely-spaced' detection sequence can thus produce at most a single sifted bit.

Rogers et al. [14] proposed a sifting algorithm with the goal of allowing the system to work in the high-speed regime without compromising the security of the key. The hope is that the key can be generated at a rate higher than $1/(2\tau)$ – the maximum key rate achievable by deactivating all detectors upon any detector firing, as we will discuss in section 4.

Rogers et al.'s algorithm can be defined as follows. Each basis is treated individually; the status of the two detectors in the other basis is irrelevant. A basis is considered active if both detectors in that basis are active at the expected photon arrival time, given by the probability $P_{0,0}$. Only in this case the subsequent detection sequence will be accepted. For an accepted detection sequence, the probability of sifting a bit from it depends on whether Alice and Bob used the same basis at least once, and ranges from 0.5 (for a single detection) up to 1 (for a very long detection sequence).

It is useful to consider a quantity $k$, the number of transmission periods per dead time, defined by $k = \rho\tau$, where $\rho$ is the transmission rate. The transition between the standard and the high-speed regimes occurs at $k = 1$, irrespective of channel loss. Rogers et al. [14] showed that the probability of a basis being active, and thus capable of sifting a bit, tends to zero as $k$ (and the transmission rate) tend to infinity. This means that high-speed QKD systems are paralyzable counting systems. This phenomenon occurs from the collective behaviour of a pair of detectors in a given basis which lock up.

Increasing the transmission rate tends to lock up the detectors in each basis, resulting in a long string of closely spaced detections which allow at most one bit to be sifted. This effect reduces and eventually outweighs the advantage of transmitting at a higher rate. The point where these balance gives rise to an optimum transmission rate that maximizes sifted bit rate production. Plotting the log of transmission rate vs sifted key rate shows a Bell-like curve. That is to say, the sifted key rate reaches a maximum value at the optimal transmission rate, after which further increases in transmission rate actually hinder sifted key production. This maximum key rate achievable with Rogers et al.'s algorithm is considerably higher than $1/(2\tau)$, given approximately by $1.43/(2\tau)$ [14].

### 3. Problems with Rogers et al.'s algorithm with active Eve

Rogers et al. paper [14] considers a purely passive Eve who can only 'listen'. Such an assumption is clearly not valid in any realistic setting where the channel is noisy and Eve can be active. We extend the analysis by introducing an Eve that can perform an intercept-resend attack.

The intercept-resend attack is a simple, yet effective, attack that involves Eve intercepting Alice's photons individually, measuring them in one of two bases used by Alice and Bob, and sending new photons to Bob according to the outcome of her measurement. Eve gains information at the cost of introducing quantum bit errors.

Eve has the ability to intercept any (or all) pulses, and resend one of Alice's states at will (also single photons), or none at all (blocking the signal). Eve is assumed to have a 4-detector set-up like Bob, but her detectors have infinitesimal dead time.

In what follows, we will show that Rogers et al.'s sifting algorithm can no longer be considered secure. This is because Eve can force Bob's detection efficiency to be basis-dependent. To illustrate this, consider a simple attack by Eve that goes as follows. Eve blocks Alice's pulses 1,2,…,$N$ (where $N$ is a large number). In their place, she sends $N$ pulses all in the "vertical" polarization state. Summing over the corresponding detection events, the probability of basis 1 (rectilinear) being active is higher than the probability of basis 2 (diagonal) being active. When $N$ is large enough we reach the stationary distribution. Although this specific attack can be easily detected by Alice and Bob in practice (because Eve's $N$ "vertical" photons will cause a high error rate), notice that Eve may lower the error rate she introduces by performing the attack on only part of the signals. Here comes a key point: Eve's ability to introduce basis-dependent detection efficiency violates a fundamental assumption in security proofs [8,20-22]. Note that for repetition rate smaller than $1/\tau$, both bases are active with equal probability.

A simulation was carried out to test for basis dependence on Bob's side in the case of this simple attack, with finite detector dead times. The probability $P_{0,0}$ of both detectors being active in each basis was calculated using Rogers et al.'s algorithm for a large number of pulses. See figure 3.

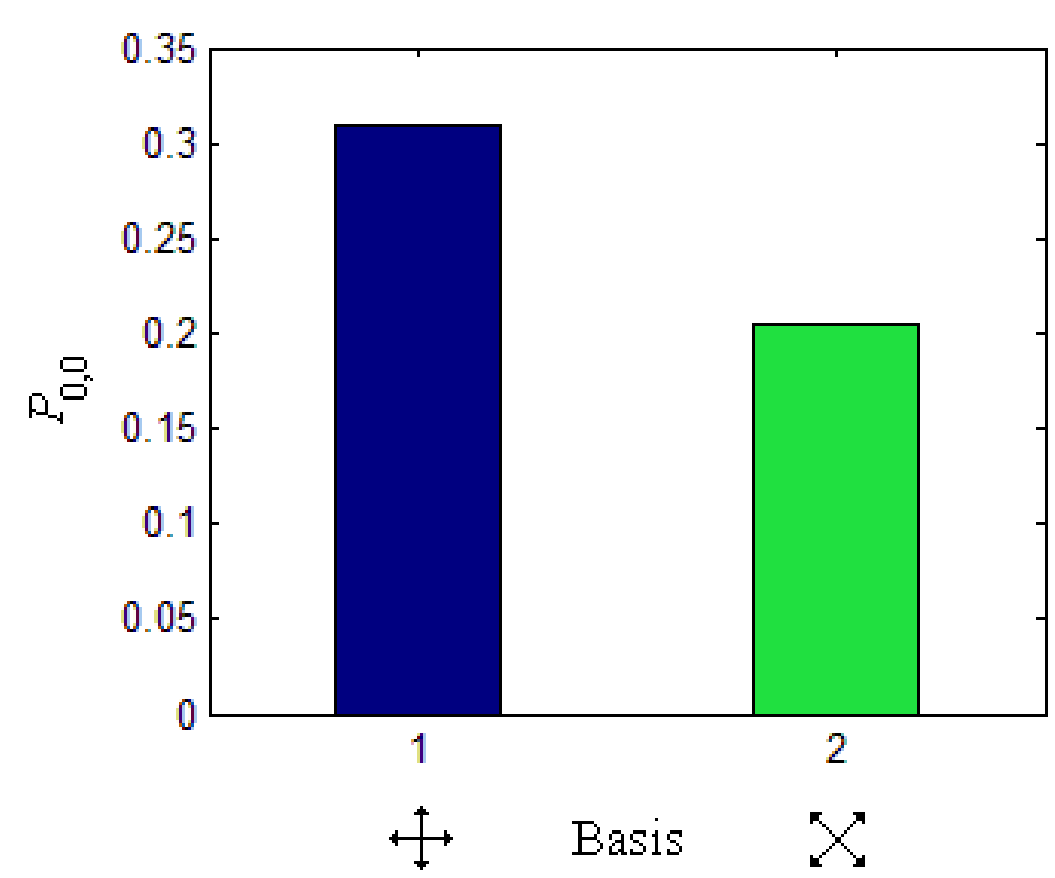

**Figure 3.** A simple intercept-resend attack can force a basis dependence of detector efficiency. Normalized transmission rate $k$ = 10, channel loss = 3 dB, detection efficiency $\eta_D$ = 100%. In the low-speed regime both bases would be active 100% of the (expected photon arrival) time.

The result is that basis one is considerably more active than basis two. Note that the ratio of $P_{0,0}$ for basis 2 to $P_{0,0}$ for basis 1 (equal to 0.67 above) is independent of the number of pulses sent as long as the number is large, and it is therefore not a transient effect caused by detectors starting off in the active state.

It is instructive to see how this basis dependence varies with *k*. We want to derive the dependence of $P_{0,0}$ on *k* for each basis. For basis 2, $P_{0,0}$ has already been derived in (8) of Ref. [14], and we quote the result here:

$$P_{0,0}(\text{basis2}) = \left[1 + (2k')\left(\frac{2p}{1-2p}\right) + (k'^2 - k')\left(\frac{(2p)^2}{1-2p}\right)\right]^{-1}, \tag{3}$$

where *p* is the probability that a particular detector produces a sifted bit on a given clock cycle and *k*' is the number of transmission periods per dead time. More concretely, for BB84 with four detectors,

$$p = \eta / 8\,, \tag{4}$$

where $\eta$ is the idealized gain defined in (1), and the factor of 1/8 accounts for the correct basis choice (1/2) and the specific detector clicking (1/4). The definition of the normalized transmission rate used by Rogers et al. [14] is slightly different than in this paper. To account for this, (3) has to be adjusted by replacing *k*′ with $k' = (k - 1)$. Note that (3) applies only at integer values of *k*′.

In Ref. [14], (3) above was derived for the case where Eve is passive and thus, in each basis, Bob receives a random bit on average. Notice that, in our case Eve always sends a vertical photon. Nonetheless, for basis 2, Bob also receives a random bit. Therefore, the derivation in Ref. [14] carries over directly here.

To find $P_{0,0}$ for basis 1 we only need to find the probability of detector 1 being active, since detector 2 is always active. The basis, at any discrete point (expected photon arrival time), is either active or passive. Let us assume for the moment that the detection system comprises only of basis 1. Then, the probability of a click given that the basis is active is given by the idealized gain for one basis $\eta_1$. Similarly, the probability of the basis remaining active is given by $1 - \eta_1$:

$$\begin{aligned} P(\text{click} \mid \text{active}) &= \eta_1 \\ P(\text{no click} \mid \text{active}) &= 1 - \eta_1 \end{aligned} \tag{5}$$

Once in a passive state, the detection system will evolve in a unique way through a series of passive states and take $(k - 1)$ steps to return to the active state. From the active state, the system at the next step can either remain active (if photon is lost), or become passive (if detector 1 is hit) and begin to recover.

We can therefore write the following equations that govern the evolution of the system in the *stationary* state, where $P_a$ is the probability that the detection system is active and $P_{p_n}$ is the probability that the system is in the passive state after *n* steps:

$$P_a = P_{p_{(k-1)}} + P_a(1 - \eta_1)$$

$$P_{p_1} = P_a \eta_1$$

$$P_{p_2} = P_{p_1} \tag{6}$$

$$\vdots$$

$$P_{p_{(k-1)}} = P_{p_{(k-2)}} .$$

We also know that the system must be in *one* of these states at any point, and since there are a total of $(k-1)$ equally likely passive states, we have

$$1 = P_a + (k-1)P_{p_1} = P_a + (k-1)P_a\eta_1 ,$$

which gives

$$P_a = \frac{1}{1+(k-1)\eta_1} . \tag{7}$$

However, there are two bases. Since only half of the photons that reach Bob will go to basis 1, we modify (7) by noting that $\eta_1 = (1/2)\eta$, where $\eta$ is the idealized gain defined by (1). This gives the probability $P_{0,0}$ that basis 1 is active (and hence capable of sifting bits):

$$P_{0,0}(\text{basis 1}) = \frac{1}{1+0.5(k-1)\eta} . \tag{8}$$

We can now see how $P_{0,0}$ changes with $k$ for the two different bases while Eve is doing the simple intercept-resend attack described above. Figure 4(a) shows both the results of Monte Carlo simulations and theoretical results (given by (3) and (8)) derived using Markov chain arguments. Note that the theoretical values apply only at integer values of $k$. Figure 3 is a snapshot of figure 4(a) for $k$ = 10. It is also interesting to see how the ratio of $P_{0,0}$ for basis 2 to $P_{0,0}$ for basis 1 varies with $k$, for the same values of $\eta$ and $p$. See figure 4(b).

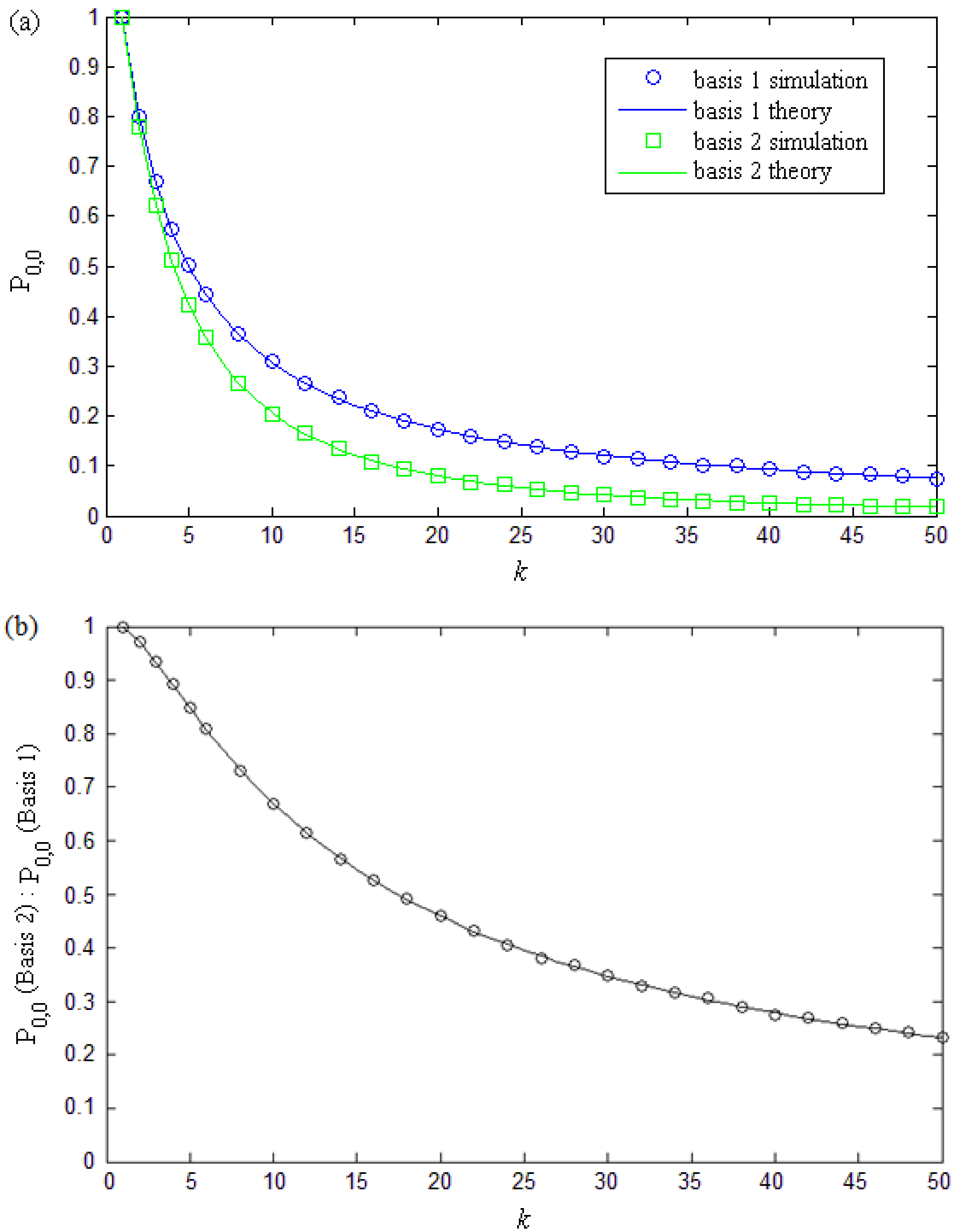

**Figure 4.** (a) Basis dependence worsens with increasing $k$. Both bases become less active, but basis 2 becomes less active at a higher rate than basis 1. $\eta$ = 0.5 (channel loss = 3 dB), so that $p = 1/16$ (as defined in (4)).
(b) Ratio of $P_{0,0}$ for basis 2 to $P_{0,0}$ for basis 1 decreases from 1 gradually to 0 with increasing $k$.

The ratio of $P_{0,0}$ for basis 2 to $P_{0,0}$ for basis 1 drops gradually to 0 with increasing $k$ from the initial value of 1 in the slow speed regime at $k = 1$.

Figures 4(a) and 4(b) clearly show how detection efficiency becomes more basis-dependent with increasing $k$. The reason for the difference between the two bases is as follows. The two detectors' dead periods in basis 2 gradually move out of sync with respect to each other after a series of independent clicks and recoveries. The chance of basis 2 being active and capable of sifting bits drops as the detectors tend to recover at different times. It becomes increasingly unlikely to have both detectors recovering at about the same time for higher values of $k$. Basis 1 on the other hand never gets locked up and *always* sifts bit value 1 when (an active) detector 1 is hit. At $k = 1$ both bases are always active. As $k$ increases, the % of time each basis is active drops as one would expect, but basis 2 suffers from the extra effect of detectors locking up, which is more prevalent for higher values of $k$.

We see that for Rogers et al.'s algorithm detection efficiency is basis dependent:

$$P_{0,0}(\text{basis } 1) \neq P_{0,0}(\text{basis } 2) .$$

This contradicts security proof assumptions [8,20-22]. Therefore, the current sifting algorithm is potentially no longer safe. Attacks attempting to exploit detection efficiency mismatch exist [23-25].

## 4. Secure sifting schemes

We have investigated alternative sifting algorithms to see if it is still possible to have secure operation for a finite dead time QKD system in the presence of the most general attack based on the dead time model we consider, assuming the original system without dead time is secure. Some of these are described below.

**AlgAllActive**
This is a purely software implementation and the algorithm only sifts a bit if all four detectors are active. It removes the aforementioned basis dependence, and the sifted bit string is secure. It is simple to implement as the detectors are free-running.

**AlgDeactivate**
In this scheme, all detectors are actively disabled (or pulses actively blocked) every time any one of the four detectors is hit [19]. This prevents any bit sifting unless all detectors are active. Again, the basis dependence is removed, and the algorithm is secure. The maximum key rate achievable is $1/(2\tau)$; the factor of $1/\tau$ comes from the maximum count rate of the individual detectors, and the factor of 1/2 comes from the fact Alice and Bob only use the right basis half the time. As this scheme involves the active disablement of all detectors, it requires an active component for its implementation.

**Alg4state**
This scheme [26,27] is different from all other considered above in that it requires only *two* detectors. Consider the phase-encoded BB84 version of this design. (This is more practical compared to the polarization-encoded version of this design as high-speed phase modulators are readily available). Bob actively selects the measurement basis for each incoming pulse. In addition, he determines which of the two detectors represents which bit value (0 or 1) for each pulse, by randomly selecting the phase modulation from a set of *four* values (0, π/2, π, 3π/2) instead of the usual two (0, π/2). A diagram representing the detection system is shown below in figure 5.

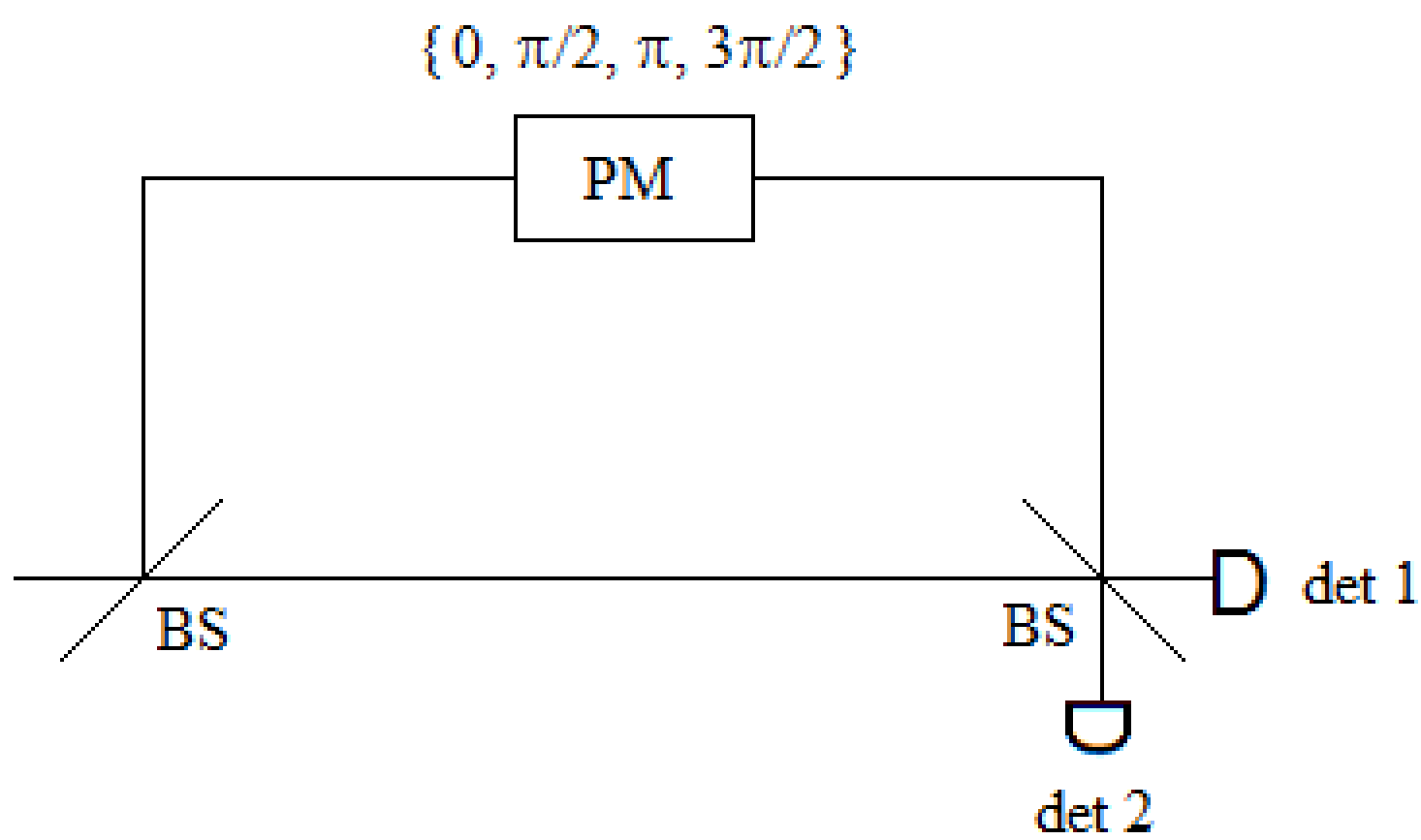

**Figure 5.** Schematic diagram of the detection system for phase-encoded version of Alg4state. BS = beamsplitter, PM = phase modulator, det 1 and det 2 = detectors 1 and 2.

*All* detections for which Alice and Bob's bases match are sifted. As such, as $k$ goes to infinity, it achieves the highest sifted key rate of all the algorithms, equal to $1/\tau$. The disadvantage is that the scheme involves more complicated modulation and requires extra random numbers.

We want to compare the sifted key rate achievable by these different algorithms. To derive the probability $P_a$ of a detection system being active for both AlgDeactivate and Alg4State, we follow the procedure outlined in section 3.

For AlgDeactivate, an active detection system constitutes all four detectors being active. The moment any one of them fires, the detection system becomes passive. $P_a$(AlgDeactivate) is therefore simply given by:

$$P_a(\text{AlgDeactivate}) = \frac{1}{1+(k-1)\eta}, \quad (9)$$

where $\eta$ is the gain, defined by the probability of a click on any of the detectors given that the detection system is active.

Only those detection events where Alice and Bob used the same bases will contribute to the sifted key, which means half of the total detection events for BB84. The sifted bit rate for AlgDeactivate is therefore given by:

$$R(\text{AlgDeactivate}) = ½\,(\text{number of clicks per second}) = ½\rho P_a(\text{AlgDeactivate})\,\text{P(click | active)}$$

$$= ½\rho P_a(\text{AlgDeactivate})\eta. \quad (10)$$

Note that the idealized gain $\eta$ is independently present in both the formula for $P_a$ in (9) and the formula for the sifted bit rate $R$ in (10). Once we have an expression for $P_a$, we still need to account for channel loss to calculate the sifted bit rate $R$. Consider the simple example where $\eta = 0$ (all signals lost), so that $P_a = 1$, and $R = 0$.

For Alg4State, since all detections in which Alice and Bob choose the same basis are sifted, it is easiest to consider the two detectors *individually*. An active detection system consists of a specific detector being active. The situation is therefore completely analogous to the derivation in section 3, and $P_a$(Alg4State) is simply given by:

$$P_a(\text{Alg4State}) = \frac{1}{1 + 0.5(k-1)\eta}. \tag{11}$$

where $\eta$ is the gain, defined by the probability of a click of a specific detector. The sifted bit rate for Alg4State is given by combining the bit rate from each of these two detectors:

$$R(\text{Alg4State}) = \tfrac{1}{2}\rho P_a(\text{Alg4State})\eta. \tag{12}$$

The factor of a half comes from three factors; 1/2 for Alice and Bob using the same basis in BB84, 1/2 to account for the photon hitting the correct one of the two detectors in an individual detection system, and finally a factor of 2 since there are effectively two independent detection systems. Again, we have to include the factor of $\eta$ to account for channel loss when calculating the sifted bit rate.

The graph below (figure 6) shows the sifted key rate as a function of the transmission rate for the different algorithms. The simulation results are based on the Monte Carlo method, while the theoretical formulas are derived above. Note that the theoretical values apply only at integer values of $k$.

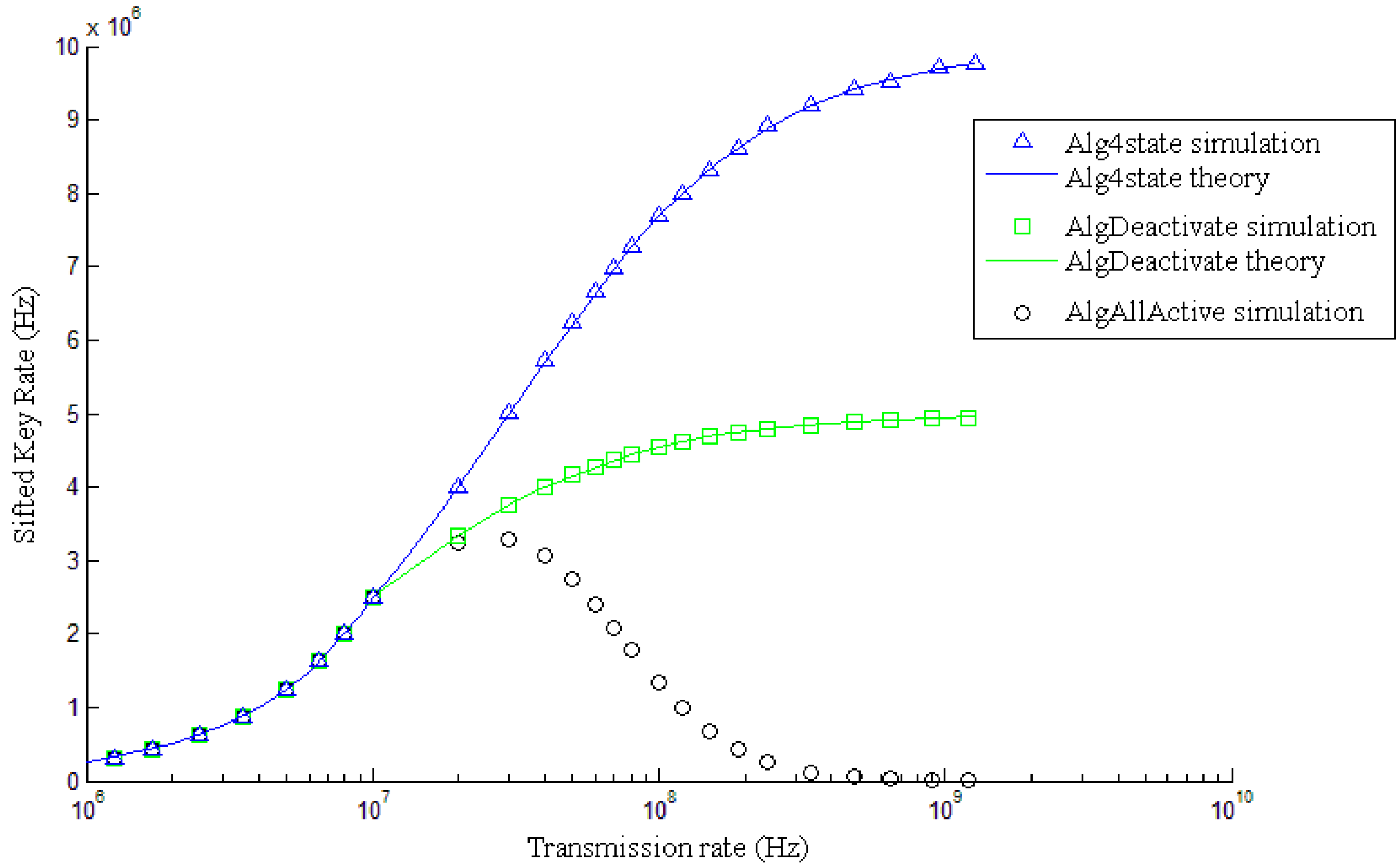

**Figure 6.** Comparison of key rates for different sifting algorithms. $\eta$ = 0.5 (channel loss = 3 dB), dead time $\tau$ = 100 ns, detection efficiency $\eta_D$ = 100%. The graph shows there is an optimum value for transmission for AlgAllActive to achieve the maximum sifted key rate. AlgDeactivate gives a higher key rate than AlgAllActive for all values of $k$ higher than 1. Alg4state gives a higher key rate than AlgDeactivate for all values of $k$ higher than 1.

The graph shows that the three algorithms are equivalent in the low-speed regime up to $k = 1$ (10 MHz on the graph since $\tau$ = 100 ns). AlgDeactivate gives a higher key rate than AlgAllActive for all values of $k$ above 1. Alg4state gives a higher key rate than AlgDeactivate for all values of $k$ above 1. The maximum key rate using AlgAllActive is achieved by transmitting in the high-speed regime (at a value of $k$ greater than 1), but not too much (exact value of $k$ depends on the dead time and channel loss).

AlgDeactivate and Alg4state do not have a peak transmission rate, but instead tend towards a constant value, given by $1/(2\tau)$ and $1/\tau$, respectively, as expected. There is no peak because in both cases the detectors do not get locked up and so the detection system is not paralyzable.

## 5. Decoy State BB84

So far we have considered a single photon source on Alice's side. Now we move onto a more realistic scenario which includes imperfections in the source and detectors.

Single photon sources are not yet practical for high speed QKD. So, weak coherent pulses (WCP) are often used. Using WCP as the source drastically reduces performance of BB84 due to multi-photon events which are susceptible to the *photon number splitting* attack [21].This has led to the development of the *decoy state* method [9-11], which allows for efficient performance even with an attenuated laser as the source.

In decoy state BB84, Alice uses more than one photon number distribution. One of these photon number distributions is optimized for the key rate. These events, called *signals*, constitute most of the pulses sent by Alice. She also sends events, called *decoys*, created with different linearly independent photon number distributions. Since Alice knows which events belong to each distribution, Alice and Bob can measure the gains (overall probability of a photon detection event for incoming pulses) for each distribution independently. This linear set of equations is used to ultimately calculate a lower bound on the key rate. Crucially, Eve must not be able to distinguish between *n*-photon events arising from different distributions.

Decoy state BB84 is now commonly used in practice after the initial experiments a few years ago [28-32].

It is interesting to analyze the security and performance of decoy state BB84 in the framework of finite detector dead time. To mitigate the effects of dead time we consider using AlgDeactivate.

In section 4, we derived the probability $P_a$(AlgDeactivate) that a detection system is active for AlgDeactivate (all four detectors are active), and hence capable of sifting bits. It is given by (9). This derivation assumes a perfect source and detectors with no false counts. We can now adjust this to account for WCP source with $\mu$ being the average photon number per pulse, and other imperfections such as background rate $Y_0$ (including dark counts and stray light).

Let us for the moment ignore the effects of dead time and consider standard decoy state protocols. We will return to the subject of dead time later. It is useful to define the following quantities [15]. The single-photon gain $Q_1$ (the joint probability that Bob's detector clicks, and that the triggering event was a single-photon) is given by

$$Q_1 = Y_1 \mu e^{-\mu}, \tag{13}$$

where $\mu$ is the average photon number per pulse and $Y_1$ is the single-photon yield (the conditional probability of a detection at Bob's side given that Alice sends a single-photon state), given by

$$Y_1 \cong Y_0 + \eta, \tag{14}$$

where $Y_0$ is the background count rate.

The overall gain $Q_\mu$ (overall detection probability summed over all individual gains) is given by

$$Q_\mu = Y_0 + 1 - e^{-\eta\mu}. \tag{15}$$

The overall quantum bit error rate (QBER) $E_\mu$ is given by

$$E_\mu = \frac{e_0 Y_0 + e_{\text{det}}(1 - e^{-\eta\mu})}{Q_\mu}, \tag{16}$$

where $e_0$ is the error of the background (taken to be 0.5) and $e_{\text{det}}$ is the probability that a photon triggered an erroneous detector (caused by e.g. optical misalignment).

The single-photon error rate $e_1$ (error rate conditioned on single-photon events) is given by

$$e_1 = \frac{e_0 Y_0 + e_{\text{det}}\eta}{Y_1}. \tag{17}$$

Let us now return to the subject of dead time. We can now modify (9) by replacing $\eta$ with $Q_\mu$, to account for WCP source and background rate $Y_0$:

$$P_{\text{a}}(\text{decoy}) = \frac{1}{1+(k-1)Q_\mu}. \tag{18}$$

Note that $P_{\text{a}}$(decoy) goes to 0 as $k$ goes to infinity. Provided that we consider the asymptotic limit of an infinitely long key and the case where the fraction of states used as decoys is negligible, we can use the following method to calculate the secure key rate of decoy BB84 with finite detector dead time:

1) Calculate the naïve key rate $R_{\text{n}}$ by assuming dead time $\tau = 0$,
2) Calculate the actual key rate $R$ by simply multiplying the naïve rate by $P_{\text{a}}$(decoy):

$$R = P_{\text{a}}(\text{decoy})R_{\text{n}}. \tag{19}$$

The naïve secure key rate $R_{\text{n}}$ (in the asymptotic limit of an infinitely long key and without dead time) is given by [15]:

$$R_{\text{n}} = q\{Q_1[1-H_2(e_1)] - Q_\mu f(E_\mu)H_2(E_\mu)\}, \tag{20}$$

where $q$ depends on the implementation (taken to be 0.5 for the BB84 protocol due to the fact that Alice and Bob use different bases half the time), $f(x)$ is the bi-directional error correction inefficiency (taken to be 1.22) and $H_2(x) = -x\log_2 x - (1-x)\log_2(1-x)$ is the binary Shannon entropy function.

The actual secure rate (per bit) with dead time is therefore given by:

$$R = P_{\text{a}}(\text{decoy})\, q\{Q_1[1-H_2(e_1)] - Q_\mu f(E_\mu)H_2(E_\mu)\}. \tag{21}$$

Together with 'GYS' parameters from an experiment in Ref. [33], we can calculate the lower bound on the actual secure key rate using (21) above. Note that in the actual GYS experiment [33] the standard BB84 protocol was used with InGaAs avalanche photodiodes operating in gated mode, while the transmission rate was just 2 MHz, so dead time effects were not significant. The value of $\mu$ is taken to be optimized for the given parameters. All the simulation parameters are summarized in table 1.

**Table 1.** Key parameters used in the simulation.

| $\tau$ (ns) | $\mu$ | $\alpha$ (dB/km) | $l$ | $e_{\text{det}}$ | $Y_0$ | $e_0$ | $\eta_{\text{Bob}}$ | $f$ |
|---|---|---|---|---|---|---|---|---|
| 100ns | 0.48 | 0.21 | 50km | 0.033 | 1.7 x $10^{-6}$ | 0.5 | 0.045 | 1.22 |

Figure 7 shows how the transmission rate affects the secure key rate both for standard decoy BB84 (without dead time effects), and decoy BB84 with dead time effects accounted for with AlgDeactivate.

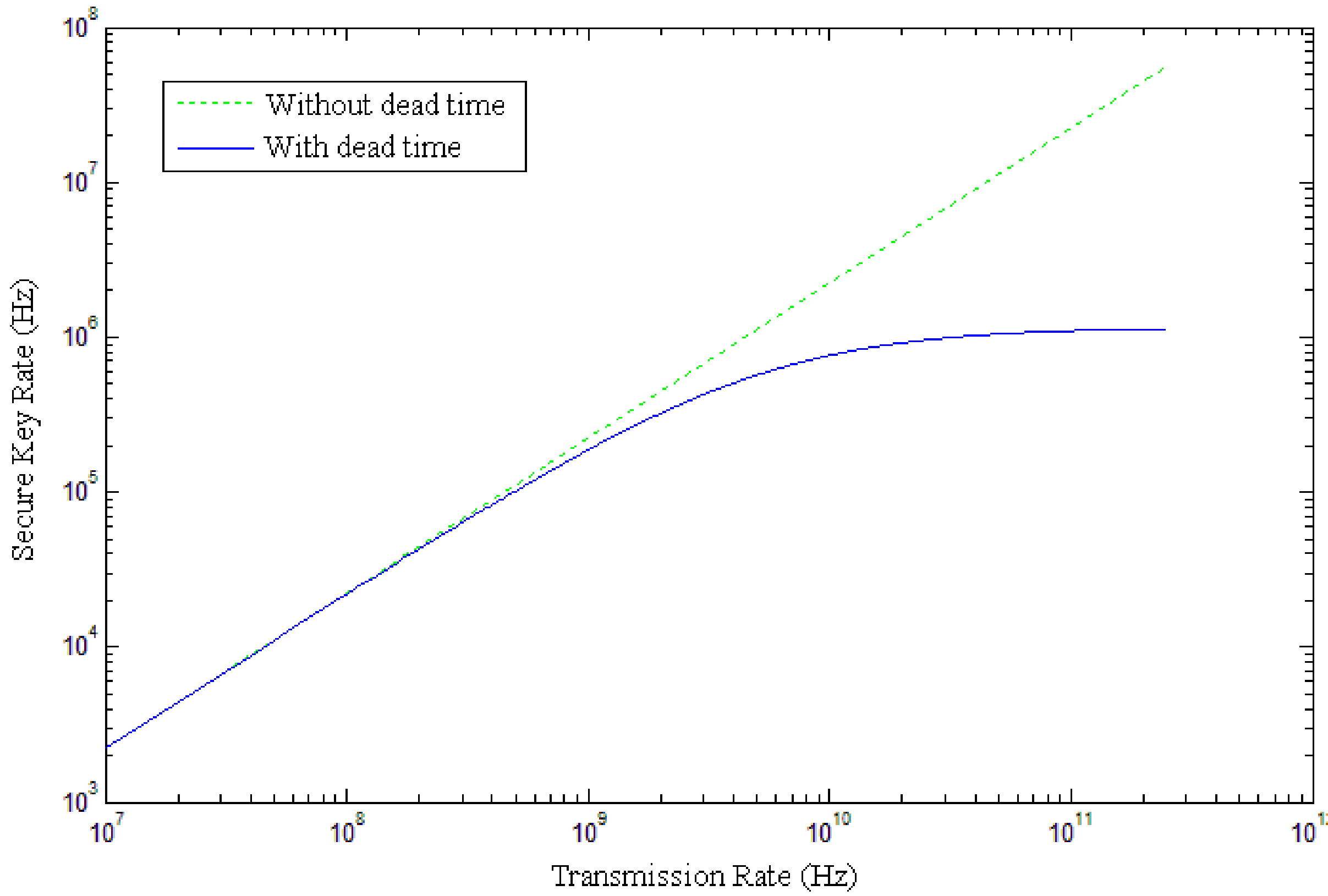

**Figure 7.** Graph showing how increasing the transmission rate affects the secure key rate. The dashed green line shows the naïve rate relation for standard decoy BB84, without any dead time effects taken into account. The solid blue line shows the actual secure key rate achievable taking dead time effects into account using AlgDeactivate.

At lower transmission rates, the two schemes yield the same secure key rate which scales linearly with transmission rate as expected. The two start to deviate as dead time effects become important. The secure key rate (per second) with the AlgDeactivate scheme (solid blue line) levels off as expected, and is given by:

$$R = \rho P_{\mathrm{a}}(\mathrm{decoy}) R_{\mathrm{n}} \, , \qquad (22)$$

where $\rho$ is the transmission rate. As $\rho$ goes to infinity, the secure key rate $R$ approaches the limiting value of $R_{\mathrm{n}}/\tau Q_{\mu}$ .

It should be noted that working with transmission rates of the order of 100 GHz might be problematic. While this high-speed phase modulation is already possible technologically (and QKD with clock rates of 10 GHz has been demonstrated [34]), the timing jitter of current detectors would be a limiting factor for the possible transmission rate. Even a small 50 ps jitter would limit the rate to approximately 10 GHz.

## 6. Conclusions

Security concerns associated with detector dead times for QKD systems operating at transmission rates higher than the maximum count rates of detectors ($1/\tau$) can limit the production rate of sifted bits. Rogers et al. [14] have proposed a sifting algorithm incorporating these dead time effects that is secure in the case where Eve is completely passive. Monte-Carlo simulations of a BB84-type QKD system with finite detector dead times were performed and extended to include an active eavesdropper, capable of interfering with the quantum channel. It was shown that the sifting algorithm proposed in Rogers et al.'s paper [14] is susceptible to intercept-resend attacks by Eve and is no longer secure because Eve is able to induce basis-dependent detector efficiency. The importance of detectors' dead periods going out of sync with each other and thus being incapable of sifting bits was highlighted. Modified sifting schemes (AlgAllActive, AlgDeactivate and Alg4State) were analyzed and compared. It was shown that a modified sifting algorithm that is secure – which sifts a bit only when all four detectors are active(AlgAllActive) – is worse in terms of maximum sifted key rate than a scheme in which all four detectors are disabled when any one of them fires (AlgDeactivate). The advantage of AlgAllActive is that it does not require any active components. The four-state protocol (Alg4state) with two detectors still achieves the highest key rate ($1/\tau$) but requires more complicated modulation and extra random numbers.

The security analysis was extended to the decoy-state BB84 protocol, and the secure key generation rate analyzed in the context of finite detector dead time.

As detectors and detection techniques improve, detector dead time is expected to be significantly reduced. Commercial Si-based products can achieve a dead time of 45 ns around 800 nm [35]. A dead time of just 1.93 ns has recently been reported with InGaAs detectors [36]. However, the dead time problem is still important to consider as the dead time could also be due to electronics in components such as the time interval analyzer (TIA) [37]. The effect of detector dead time on the security of the differential phase shift protocol [38] has also been analyzed [39].

Future directions could include a more realistic detector dead time model, including a gradual recovery to maximum efficiency, dark counts and afterpulsing. Ignoring double click events, security can still be proved for a gradual recovery [40] with, for example, Alg4state. The dead time of different detectors would also not be the same in practice. Eve's ability to control the dead time to her advantage could open up new avenues for attack.

The important lesson here is the need to check carefully the operations of components of a QKD system (in this case the detection system) and make sure that it is properly described by the security model.

**Acknowledgements**

Support of the funding agencies CIFAR, CIPI, Connaught, CRC, MITACS, NSERC, QuantumWorks and the University of Toronto is gratefully acknowledged.